\documentclass[twocolumn,secnumarabic,amssymb, nobibnotes, aps, prl]{revtex4-2}

\usepackage{graphicx}
\usepackage{subfigure}
\usepackage{bm}
\usepackage{amsmath}
\usepackage{color}
\begin{document}
	
	
	\title{Separation of bi-dispersed microspheres in dusty plasma ratchet experiments}
	
	
	
	\author{Ting-yu Yao$^{1}$}
	\author{Ji-xu Gao$^{1}$}
	\author{Miao Tian$^{1}$}
	\author{Shun-xin Zhang$^{1}$}
	\author{Fu-cheng Liu$^{1}$}
	\author{Bao-quan Ai$^{2}$}
	\email[Email:]{aibq@scnu.edu.cn}
	\author{Yan Feng$^{3}$}
	\email[Email:]{fengyan@suda.edu.cn}
	\author{Ya-feng He$^{1}$}
	\email[Email:]{heyf@hbu.edu.cn}
	
	\affiliation{$^1$College of Physics Science and Technology, Hebei Research Center of the Basic Discipline for Computational Physics, Hebei University, Baoding 071002, China\\
		$^2$Key Laboratory of Atomic and Subatomic Structure and Quantum Control (Ministry of Education), Guangdong Basic Research Center of Excellence for Structure and Fundamental Interactions of Matter, School of Physics, South China Normal University, Guangzhou 510006, China\\
		$^3$Institute of Plasma Physics and Technology, Jiangsu Key Laboratory of Frontier Material Physics and Devices, School of Physical Science and Technology, Soochow University, Suzhou 215006, China.}
	
	
	\date{\today}
	
	\begin{abstract}
		It is demonstrated experimentally that the effective separation of bi-dispersed microspheres (dust particles) in the underdamped and strongly-coupled regime is realized using a designed dusty plasma ratchet. Experimental findings reveal that these dust particles can undergo directional transport at varying speeds, even moving in opposite directions depending on the discharge conditions, enabling successful particle separation. Numerical simulations of the plasma environment surrounding the dust particles are performed using fluid simulations of the capacitively coupled discharge of Argon. The simulation results indicate that the bi-dispersed dust particles are suspended at different balance heights within the plasma sheath and experience distinct ratchet potentials that govern their directional transport, resulting in varied flow velocities. The discovery of height-dependent transport of dust particles here provides insights of transport fundamental of underdamped strongly-coupled particles in dusty plasma ratchets.
	\end{abstract}
	
	
	
	\maketitle
	
	\subsection{I. INTRODUCTION}
	\indent The fundamental mechanism of transport under various conditions, such as the particle size, the particle shape, and the external field/gradient, attracts great attention in various fields~\cite{Zhang, Mage,  Stogin, Chen, Aritra,Tang}. By manipulating the transport properties of particles, a collection of particles with different sizes may be separated or sorted effectively, which has numerous applications \cite{Skaug, Garg, Zeming}.
	
	\indent A variety of Feynman ratchets, composed of periodic and asymmetry arranged units providing the essential spatial symmetry breaking, are proposed to control the particle transport \cite{Mahmud, Arzola, Skaug}. The Feynman ratchet actually can be regarded as a heat engine which converts the random motion into the directional transport of particles without violating the second law of thermodynamics \cite{Feynman, hanggi}. Previous Feynman ratchet studies clearly indicate that the particle size is an important factor in controlling the transport velocity of particles, so that the particle separation may be achieved in Feynman ratchets \cite{Aritra,Tang,Skaug}. From our literature search of Feynman ratchets \cite{Nicollier, Schwemmer, Matthias, Lout,Motegi,Verleger}, typically the neutral or slightly charged particles are immersed in electrolytes, and extensive efforts are devoted to control the transport of these overdamped particles to separate them.
	
	\indent Here, we demonstrate the separation of underdamped strongly-coupled bi-dispersed microspheres in a dusty plasma ratchet experiment. Dusty plasma consists of microspheres, i.e., micron-size dust particles, immersed in a plasma environment \cite{Morfill,Shukla,POPREV,Chu,Wang,Thomas,Ott,Killer,Du,Feng,Bajaj,Joshi,HuangD}. In the typical laboratory conditions, these dust particles are negatively charged to the order of 10$^3$ to 10$^5 e$~\cite{Morfill}, leading to the strong coupling effect between them \cite{Wong, Kalman}. These unique features enable the dust particles to achieve rapid transport within the gaseous plasma, a feat that was previously challenging in experimental systems involving electrolytes.\\	
	\indent In~\cite{He}, it is demonstrated that, in a dusty plasma ratchet, the persistent flow of monodispersed dust particles is controlled by the gas pressure and the plasma power, where the asymmetric electric potential and the collective effect play the key roles in this flow rectification. In~\cite{Wang1}, experimental validation shown that the suspended height of monodispersed dust particles in the plasma sheath determines the flow direction. In principle, bi-dispersed particles of various sizes are levitated at different heights in the plasma sheath, so that they undergo different ratchet potentials that govern their flow velocities. By adjusting the plasma conditions, the ratchet potentials for bi-dispersed dust particles may be modified to obtain desired different flow velocities for dust particles with different sizes, leading to the corresponding bi-dispersed dust particle separation, as we study in experiments here.
	
	\begin{figure}[htp]
		\begin{center}
			\includegraphics[width=8.5cm,height=3.9cm]{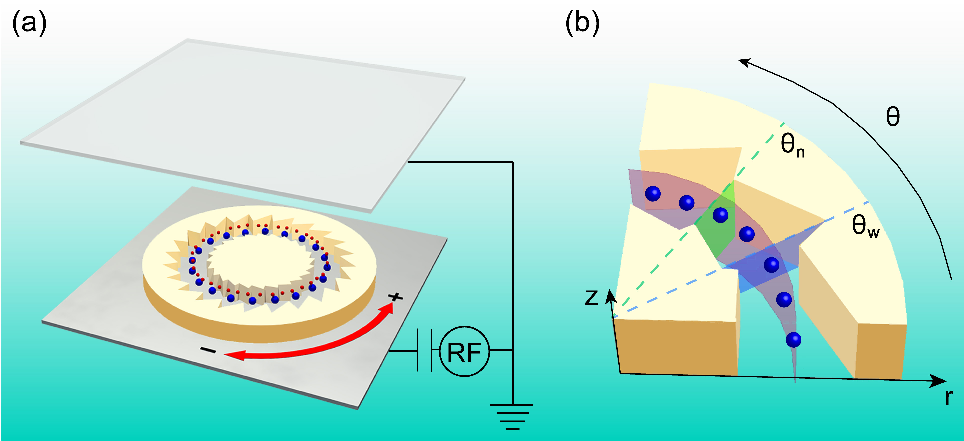}
			\caption{Diagram of the dusty plasma ratchet. (a): Bi-dispersed dust particles (red small/blue large) confined in the sawtooth channel between two gears and suspended at slightly different heights above the lower electrode. At an appropriate gas pressure and the rf power, different sized dust particles are transported at different speeds along the sawtooth channel, even in different direction, as indicated by the `+' and `$-$' symbols. (b): Enlarged 3D view of the sawtooth channel. The cylindrical surface along the center of the saw channel indicates the $z$-$\theta$ frame used in Figs.~4 and 5.}
		\end{center}
	\end{figure}
	
	\subsection{II. DUSTY PLASMA RATCHET}
	\indent The experimental setup is the same as that used in~\cite{He}. The dusty plasma ratchet used here consists of two resin gears with asymmetrical sawtooth placed concentrically on a horizontal lower electrode, as shown in Fig.~1. The dimensions of the inner and outer gears are 9 mm in height, 11.75 and 23 mm in the average radius, 1.5 and 4 mm in the depth, respectively. The two gears have the same orientation and sawtooth number, so that a circular sawtooth channel is generated between two gears. The width of the enclosed sawtooth channel changes periodically and asymmetrically in the azimuthal ($\theta$-) direction. The distance between the upper and lower electrodes is 4.5 cm. During the experiment, plasma is generated using the Argon gas at the pressure $p$ varying from 15 to 40 Pa between two electrodes and ignited by the capacitively coupled 13.56 MHz radio frequency (rf) source with the power $P$ varying from around $<$ 10 W to 40 W, measured from the rf amplifier. Plasma sheath is generated above the surface of the gears and the electrode, with the profile roughly following the shape of the sawtooth channel. When macroporous cross-linked polystyrene microspheres (dust particles) are introduced into the sawtooth channel, they are charged immediately and suspended around the center of the sawtooth channel due to the confinement from the electric field of sheath. Under suitable conditions, persistent flows of dust particles are generated along the sawtooth channel. Here, we choose two sizes of dust particles with the radii of $r_d$ = 8 and 14 $\mu$m, respectively, so that they are levitated at slightly different heights in the plasma sheath. During experiments, dust particles are illuminated using a flood lamp positioned on one side of the chamber, so that their motion is recorded using a top-view camera operating at 50 frames per second. For better particle identification and tracking, we set the particle number to be $\sim$ 200 for each particle size, leading to the interparticle distance of $\sim$ 0.5 mm in the dust chain.
	
	\begin{figure}[htbp]
		\begin{center}\includegraphics[width=8.5cm,height=6.9cm]{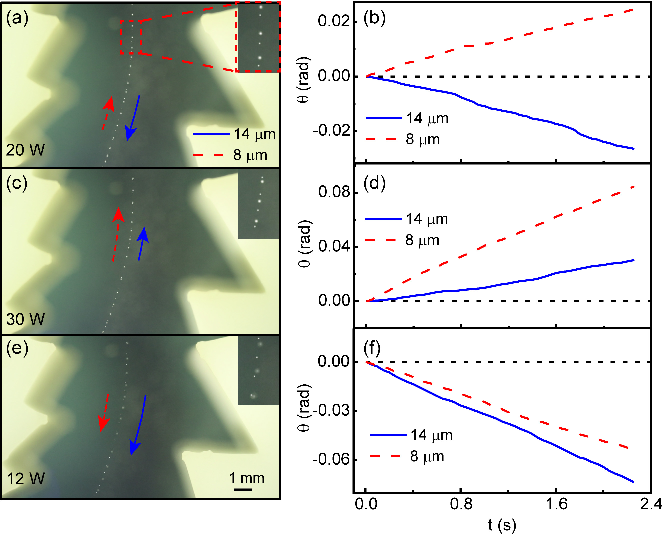}
			\caption{Separation of bi-dispersed dust particles from our experiments. Two sized dust particles with the radii $r_d$=8 and 14 $\mu$m are transported in the positive and negative directions, respectively, at 25 Pa, 20 W, as indicated by the arrows in (a) and the angular variation $\theta$-$t$ in (b). When the conditions are changed to 25 Pa and 30 W, the two sized dust particles are both transported along the positive direction, while at two different angular speeds, as shown in (c) and (d). When the conditions are 25 Pa and 12 W, the two sized dust particles are transported along the negative direction at two different angular speeds, as shown in (e) and (f).}
		\end{center}
	\end{figure}
	
	\subsection{III. PARTICLE SEPARATION EXPERIMENTS}
	\indent As the major result of this paper, we realize different persistent dust flows with two distinctive velocities in the transport of the bi-dispersed dust particles in our dusty plasma ratchet experiment, as presented in Fig.~2. The velocities of these two dust flows can be controlled conveniently by adjusting the plasma discharge conditions. Figure 2(a) clearly shows an unambiguous separation of the bi-dispersed dust particles in the two opposite directions, under the conditions of $p$ = 25 Pa and $P$ = 20 W. The azimuthal locations of two selected dust particles with the different radii, presented in Fig.~2(b), clearly illustrate the effective separation of the bi-dispersed dust particles moving in the opposite directions.
	
	\indent In fact, even flowing in the same directions, the separation of these bi-dispersed dust particles are also achieved when their moving speeds are significantly different, as shown in Figs.~2(c)-2(f). When the rf power is changed to $P$ = 30 W, both the smaller and larger dust particles flow in the positive directions, but at different speeds as indicated by their respective trajectory lengths during 2.3 seconds in Fig.~2(c). From the variation of their azimuthal locations in Fig.~2(d), the calculated angular speed of the smaller dust particle $\sim$ 0.038 rad/s is about 3 times larger than that for the larger dust particle $\sim$ 0.013 rad/s, further confirming the separation of these bi-dispersed dust particles. When the rf power is changed to $P$ = 12 W, two sized dust particles flow in the negative directions with different speeds are also achieved, as shown in Figs.~2(e), 2(f). The particle motion movies under these conditions are provided in the Supplementary Video \cite{SM}.
	
	\begin{figure}[htbp]
		\begin{center}\includegraphics[width=8cm,height=6.4cm]{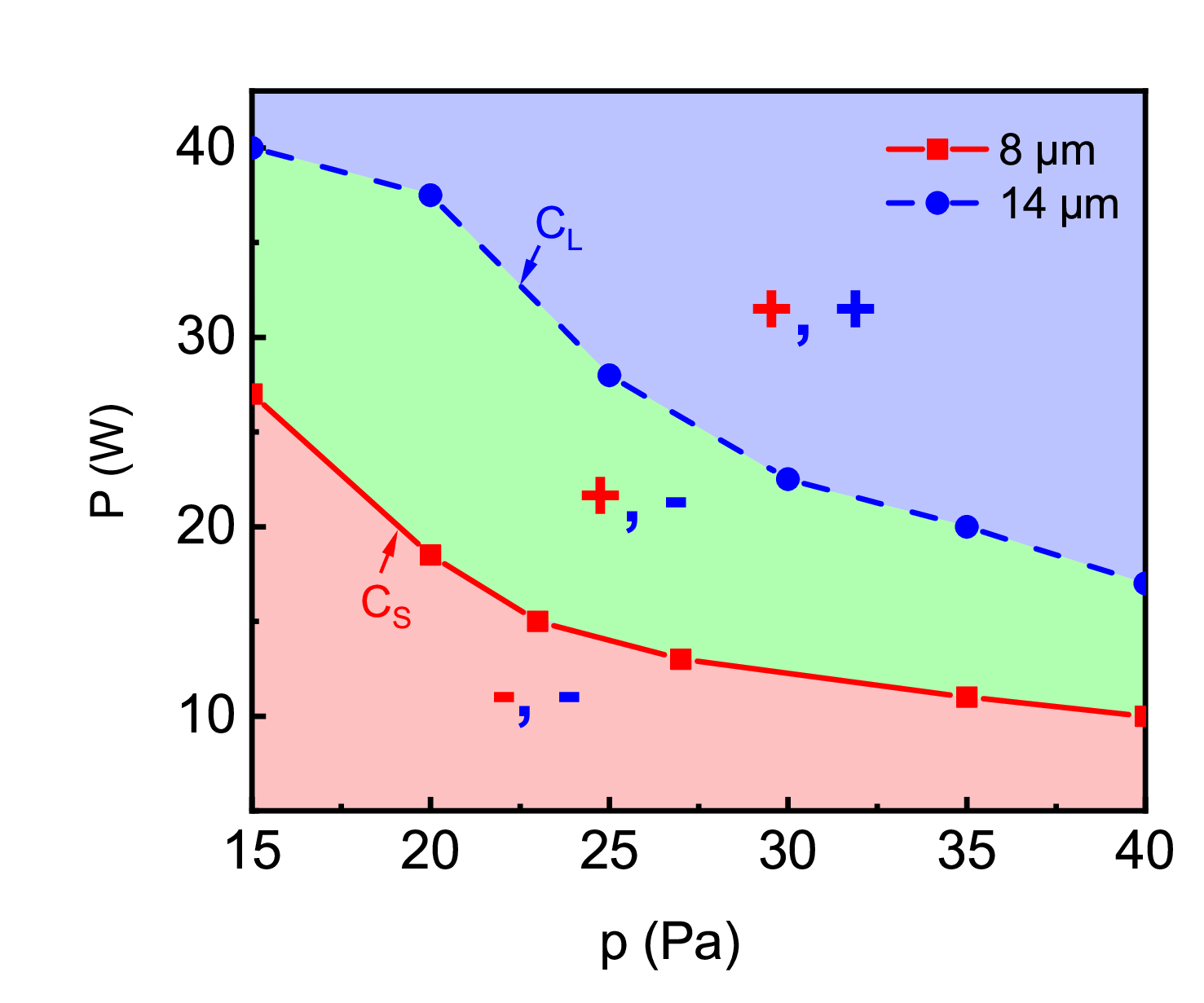}
			\caption{Phase diagram of particle separation under varying gas pressure and power from our experiments. The solid critical curve $C_{S}$ separates the phase diagram into positive `+' and negative `$-$' flow regions of the directional transport for the dust particles with the size of $r_d$ = 8 $\mu$m. The dashed critical curve $C_{L}$ indicates the case for $r_d$ = 14 $\mu$m. In the middle `+$-$' region, the two sizes of dust particles flow in opposite directions. In the `++' (`$-$$-$' ) region, the bi-dispersed dust particles flow in the same positive (negative) directions.}
		\end{center}
	\end{figure}
	
	\indent The dusty plasma ratchet can be operated in a wide range of the plasma conditions for the separation of bi-dispersed particles, since the transport of dust particles is easily controlled by changing the plasma conditions. Our experimentally obtained ``phase'' diagram for the particle motion direction for dust particles with the two radii of $r_d$ = 8 and 14 $\mu$m is presented in Fig.~3. The two critical curves of $C_S$ and $C_L$ divide this diagram into the positive `+' and negative `$-$' regions of the directional flow of the dust particles with the size of $r_d$ = 8 $\mu$m and 14 $\mu$m, respectively. In fact, as confirmed by our test experiments, under any conditions in the central `+$-$' region, the smaller dust particles always flow in the positive direction, while the larger dust particles are transported reversely, as the example presented in Fig.~2(a). For the conditions in the `++' or `$-$$-$' regions of Fig.~3, eventhough they flow in the same direction, the separation of the bi-dispersed dust particles can also be achieved by the different flow speeds as well, as the examples presented in Figs.~2(c) and 2(e).
	
	\subsection{IV. MECHANISM OF PARTICLE SEPARATION}
	\indent We attribute the observed separation phenomena to the different responses of the bi-dispersed dust particles to the plasma conditions during their directional transports. The plasma parameters surrounding dust particles determine the charging currents of ions and electrons, the electric force balancing their gravity, and the ion drag force on the dust particles, influencing their transports. For bi-dispersed dust particles, they are suspended at different heights within the sawtooth channel and consequently experience distinctive plasma environments, leading to their different transport behaviors. 
	
	\indent {We perform numerical simulations to calculate 3D distribution of plasma parameters within the narrow sawtooth channel based on the dusty plasma ratchet model using two steps, as described in detail in \cite{He, Wang1}. First, utilizing 2D fluid simulations of the capacitively coupled discharge of Argon plasma through COMSOL Multiphysics software \cite{COMSOL}, we calculate distribution of the plasma parameters $\Pi (r,z)$ in two typical sections ($z$-$r$ frame), corresponding to the widest and narrowest vertical sections of the sawtooth channel, indicated in Fig.~1(b). Second, for any points between these two vertical sections, we perform interpolation between the two vertical sections in term of a double-sine function describing the change of plasma parameters $\Pi (\theta)$ with the azimuthal position $\theta$. As a result, we obtain the full 3D distribution of plasma parameters $\Pi (r,z,\theta)$ within the sawtooth channel, very similar to that in~\cite{He}. 
	
	Note, in our experiments, both the smaller and larger dust particles are radially confined to the center ($r$$\approx$14.3 mm) of the sawtooth channel, as shown in Fig.~2. Thus, the horizontal transport (in the $\theta$- direction) of the bi-dispersed dust particles suspended at different heights ($z$) are actually restricted only on a 2D cylindrical surface ($z$-$\theta$ frame at $r$$\approx$14.3 mm) as denoted in Fig.~1(b). Therefore, we just focus here on the distribution of numerically calculated plasma parameters and the directional transport of the dust particles on this cylindrical surface.
	
	\begin{figure}[htp]
		\begin{center}\includegraphics[width=8cm,height=7.4cm]{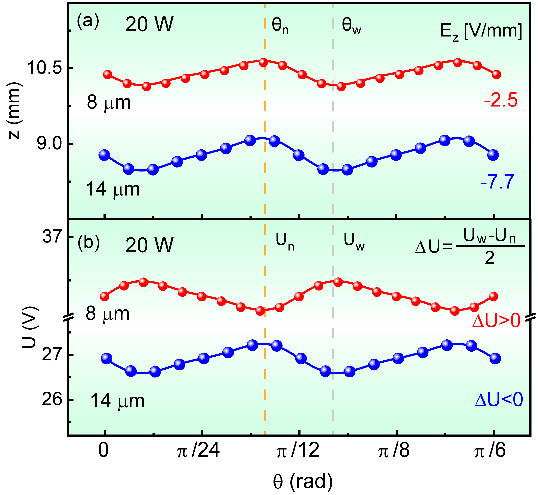}
			\caption{Calculated suspended heights (a) and experienced ratchet potentials (b) of bi-dispersed particles at the discharge power $P$ = 20 W from our simulations. Dust particles with identical sizes are levitated along a contour of the vertical electric field $E_z$, and consequently undergo the action of the ratchet potential. Bi-dispersed dust particles (8 $\mu$m, red balls and 14 $\mu$m, blue balls) suspended at different heights undergo the action of the corresponding ratchet potentials with opposite orientations, resulting in the particle flow in opposite directions along the sawtooth channel. Here, $U_w$ and $U_n$ represent the extremal potentials at the azimuthal positions $\theta_w$ and $\theta_n$, respectively. Gas pressure $p$ $=$ 25 Pa.}
		\end{center}
	\end{figure}
	
	\indent We first examine the mechanism behind the bi-directional separation of bi-dispersed particles at $P$ = 20 W, as illustrated in Fig.~4. For a dust chain composed of identical dust particles, its suspended height on the cylindrical surface would follow a contour of the vertical electric field $E_z$ = $-\partial U/\partial z$, as shown in Fig.~4(a). Clearly, the suspension height profile exhibits asymmetric wavy curves in the $\theta$- direction due to the periodic and asymmetric changes in the width of the sawtooth channel. Here, the suspended height of charged dust particles is primarily determined by the vertical electric field $E_z$, which balance the gravity, $mg$ = $QE_z$, since the charges $Q$ of dust particles remains nearly constant while flowing along the sawtooth channel, as confirmed in [35]. Obviously, for the dust particles with the sizes of 8 $\mu$m and 14 $\mu$m, these particles are levitated at different heights along two distinctive contours of $E_z$, as indicated in Fig.~4(a).
	
	\indent A dust chain composed of identical dust particles experiences the influence of the ratchet potential along the azimuthal ($\theta$) direction. From our numerically calculated results presented in Fig.~4(b), the electric potential at the suspended height of dust particles varies along the $\theta$ direction. This electric potential exhibits both asymmetric and periodic characteristics, corresponding to the form of a ratchet potential, which can be depicted by a double-sine function \cite{He}:
	\begin{eqnarray}
		U(\theta)&=&\triangle U\Big(\sin(\frac{2\pi\theta}{\Theta})+\frac{1}{4}\sin(\frac{4\pi\theta}{\Theta})\Big)+U_0.
	\end{eqnarray}
	Here, the sign and magnitude of $\triangle U$ characterize the asymmetrical orientation and barrier height of the ratchet potential, respectively. For the bi-dispersed dust chains suspended at different heights, the ratchet potentials they experience have significantly different characteristics. Our calculated results show that at the discharge power of $P$ $=$ 20 W, the bi-dispersed dust chains are subjected to ratchet potentials with completely opposite orientations of asymmetry, as depicted in Fig.~4(b). 

	\indent The encountered ratchet potential governs the directional transport of the dust chain along the sawtooth channel. The ratchet potential $U(\theta)$ induces an azimuthal electric field $E_{\theta}$ which consequently causes an azimuthal ion flow. This ion flow further applies the ion drag force $F_{i\theta}$ to the dust particles. Although the signs of the azimuthal electric field and the ion drag force are opposite on the two side of the ratchet potential well, due to the asymmetry of the ratchet potential, the integral $f_{i\theta}$ = $\int_{0}^{2\pi}{F_{i\theta}}d{\theta}$ along the whole sawtooth channel could be nonzero, which is able to drive the dust chain to flow along the sawtooth channel. The sign of the integral $f_{i\theta}$ correlates with the asymmetric orientation of the ratchet potential (sign of $\triangle U$), which determines the flow direction of the dust chain. Meanwhile, the magnitude of $f_{i\theta}$ is influenced by the strength of $E_{\theta}$ (which is associated the magnitude of $\triangle U$), thereby affecting the flow velocity of the dust chain. Consequently, at the discharge power $P$ $=$ 20 W, the bi-dispersed particles undergo ratchet potentials with opposite orientations and flow in the opposite directions along the sawtooth channel. 
	
	\begin{figure}[htp]
		\begin{center}\includegraphics[width=8cm,height=9.7cm]{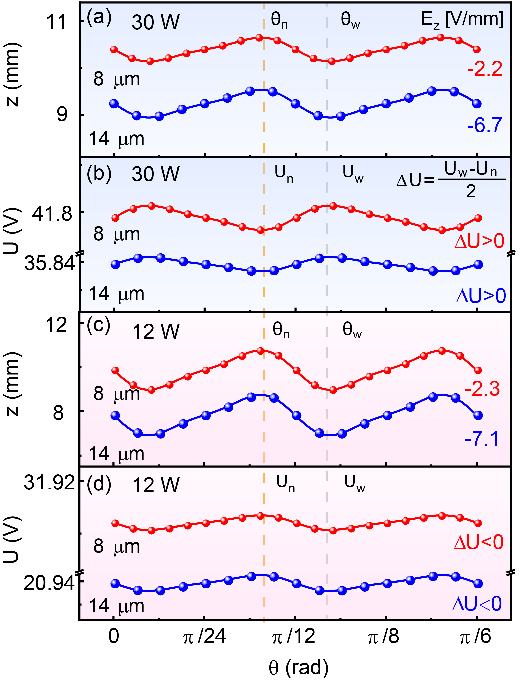}
			\caption{Calculated suspended heights and experienced ratchet potentials of bi-dispersed particles at the discharge powers (a)-(b) $P$ = 30 W, (c)-(d) $P$ = 12 W from our simulations. For $P$ = 30 W, both small and large dust particles experience ratchet potentials with the same positive orientation but different barrier heights, resulting in their positive flow along the sawtooth channel, but different velocities. For $P$ = 12 W, the opposite is true. Gas pressure $p$ $=$ 25 Pa.}
		\end{center}
	\end{figure}

	\indent As the discharge power is varied, the suspended heights of the bi-dispersed particles change accordingly, leading to significant alterations in the experienced ratchet potentials, as illustrated in Fig.~5 from simulations. When the discharge power is increased to $P$ $=$ 30 W, the suspended heights of the bi-dispersed particles rise to a higher level. In this scenario, they experience ratchet potentials with the same positive orientation ($\triangle U$ $>$ 0) but different barrier heights, as depicted in Figs.~5(a) and 5(b). This results in their positive flow along the sawtooth channel, with distinct velocities. Conversely, when the discharge power is decreased to $P$ $=$ 12 W, the suspended heights of the bi-dispersed particles lower. In this case, they undergo ratchet potentials with the same negative orientation ($\triangle U$ $<$ 0) but still with varying barrier heights, as shown in Figs.~5(c) and 5(d), leading to a negative flow with different velocities. In summary, by carefully controlling the discharge power to manipulate the suspended heights of the bi-dispersed particles and the associated ratchet potentials, effective particle separation is naturally achieved.
	
	\begin{figure}[htp]
		\begin{center}\includegraphics[width=8cm,height=10.1cm]{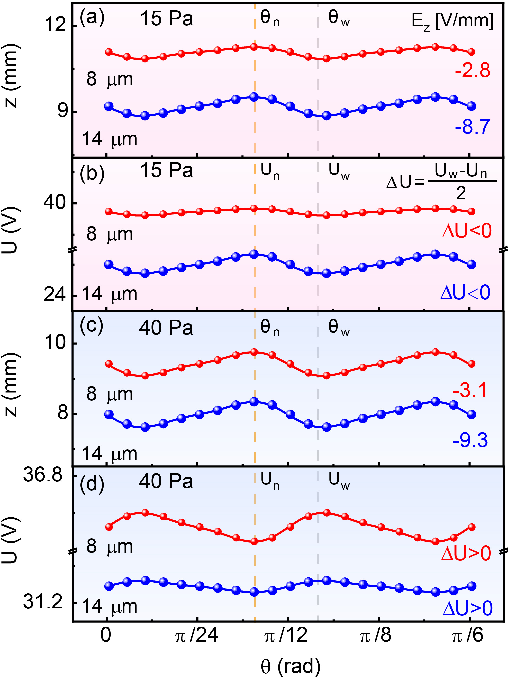}
			\caption{Calculated suspended heights and experienced ratchet potentials of bi-dispersed particles at the gas pressures (a)-(b) $p$ = 15 Pa, (c)-(d) $p$ = 40 Pa from our simulations. For $p$ = 15 Pa, both small and large dust particles experience ratchet potentials with the same negative orientation but different barrier heights, resulting in their negative flow along the sawtooth channel, but different velocities. For $p$ = 40 Pa, the opposite is true. Discharge power $P$ $=$ 20 W.}
		\end{center}
	\end{figure}
	
	\indent As the gas pressure changes, the suspended heights of the bi-dispersed particles shift due to corresponding variations in the sheath thickness, also leading to significant changes in the ratchet potentials that dust particles experience, as illustrated in Fig.~6 from our simulations. As compared to Fig.~4, when the gas pressure is reduced to $p$ = 15 Pa at a fixed discharge power of $P$ = 20 W, the bi-dispersed particles experience ratchet potentials with the same negative asymmetry orientation but different barrier heights. This configuration drives both particle types to flow in the same negative direction, but at different velocities, as shown in Figs.~6(a) and 6(b). However, when the gas pressure is increased to $p$ = 40 Pa, the ratchet potentials of both particle types exhibit the same positive asymmetry, driving these particles move in the positive direction along the sawtooth channel, also at different velocities, as presented in Figs.~6(c) and 6(d).
	
	
	\subsection{V. DISCUSSION AND CONCLUSION}
	\indent From extensive experiments with different particle sizes ranging from 5 to 16 $\mu$m, we conclude that as long as the size difference of the bi-dispersed dust particles is larger than $\sim$ 2 $\mu$m, they could be separated in opposite directions under appropriate discharge conditions. If the size difference is too small, the ion wake effect from the upper small dust particles would influence the motion of the lower large dust particles, diminishing the separation efficiency. The flow speeds of the underdamped dust particles in the dusty plasma ratchet can reach $\sim$ 1 cm/s as we tested upon extending the experimental conditions to wider ranges, suggesting a better separation performance.
	
	\indent Our results of the separation of bi-dispersed dust particles under fixed plasma conditions significantly exceeds the previous experimental results focusing on the rectification of monodispersed dust particles under varying plasma conditions \cite{He, Wang1}. Although the experiment in \cite{Wang1} involved employing two sizes of dust particles, its purpose is to utilize the ion wake from the smaller dust particles to lift the suspended height of the studied larger dust particles in order to change the directional transport of the larger dust particles. However, in the present separation experiments, the effect of the ion wake must be avoided to achieve perfect separation efficiency. In addition, based on our current results of particle separation in the circular ratchet structure in Argon plasma, a straight ratchet structure with two outlets to collect dust particles with different sizes is also successfully designed \cite{Tian}, so that the separation of bi-dispersed dust particles in Air plasma is feasible in different ratchet scenarios.
	
	\indent We demonstrate the transport and separation of bi-dispersed dust particles using a designed dusty plasma ratchet. Effective separation of the underdamped dust particles is easily achieved in this dusty plasma ratchet. Simulated results illustrate that different sizes of dust particles are levitated at distinct balance heights and undergo dissimilar ratchet potentials, resulting in their different flow velocities. This ratchet-based experimental method offers new insight into the separation of underdamped strongly-coupled particles and could be used for developing advanced particle separation devices.
	
	\subsection{ACKNOWLEDGMENTS}
	\indent This work is supported by the National Natural Science Foundation of China (Grant Nos. 12275064, 12475203, 12475036, 12075090), the Natural Science Foundation of Hebei Province of China (Grant No. A2024201020), the Scientific Research and Innovation Team project of Hebei University (Grant No. IT2023B03). Work in Suzhou is supported by the National Natural Science Foundation of China under Grant No. 11875199 and the 1000 Youth Talents Plan.


\begin{thebibliography}{}
		
		\bibitem{Zhang} J. Zhang, R. Alert, J. Yan, N. S. Wingreen, and S. Granick, Active phase separation by turning towards regions of higher density, Nat. Phys. $\mathbf{17}$, 961-967 (2021).
		\bibitem{Mage} P. L. Mage, A. T. Csordas, T. Brown, D. Klinger, M. Eisenstein, S. Mitragotri, C. Hawker, and H. T. Soh, Shape-based separation of synthetic microparticles. Nat. Mater. $\mathbf{18}$, 82-89 (2019).
		\bibitem{Stogin} B. B. Stogin, L. Gockowski, H. Feldstein, H. Claure, J. Wang, and T. S. Wong, Free-standing liquid membranes as unusual particle separators, Sci. Adv. $\mathbf{4}$, eaat3276 (2018).
		\bibitem{Chen} R. T. Chen, S. Pang, H. Y. An, J. Zhu, S. Ye, Y. Y. Gao, F. T. Fan, and C. Li, Charge separation via asymmetric illumination in photocatalytic Cu$_2$O particles, Nat. Energy $\mathbf{3}$, 655-663 (2018).
		\bibitem{Aritra} A. K. Mukhopadhyay, B. Liebchen, and P. Schmelcher, Simultaneous control of multispecies particle transport and segregation in driven lattices, Phys. Rev. Lett. $\mathbf{120}$, 218002 (2018).
		
		\bibitem{Tang} H. Tang, J. Q. Niu, H. Jin,  S. J. Lin, and  D. X. Cui, Geometric structure design of passive label-free microfluidic systems for biological micro-object separation, Microsyst. Nanoeng. $\mathbf{8}$, 62 (2022).
		\bibitem{Skaug} M. J. Skaug, C. Schwemmer, S. Fringes, C. D. Rawlings, and A. W. Knoll, Nanofluidic rocking brownian motors, Science $\mathbf{359}$, 1505-1508 (2018).
		\bibitem{Garg} N. Garg, T. M. Westerhof, V. Liu, R. Liu, E. L. Nelson, and A. P. Lee, Whole-blood sorting, enrichment and in situ immunolabeling of cellular subsets using acoustic microstreaming, Microsyst. Nanoeng. $\mathbf{4}$, 17085 (2018).
		\bibitem{Zeming} K. K. Zeming, S. Ranjan, and Y. Zhang, Rotational separation of non-spherical bioparticles using I-shaped pillar arrays in a microfluidic device, Nat. Commun. $\mathbf{4}$, 1625 (2013).
		\bibitem{Mahmud} G. Mahmud, C. J. Campbell, K. J. M. Bishop, Y. A. Komarova, O. Chaga, S. Soh, S. Huda, K. Kandere-Grzybowska, and B. A. Grzybowski, Directing cell motions on micropatterned ratchets, Nat. Phys. $\mathbf{5}$, 606-612 (2009).
		
		\bibitem{Arzola} A. V. Arzola, M. Villasante-Barahona, K. Volke-Sep\'{u}lveda, P. J\'{a}kl, and P. Zem\'{a}nek, Omnidirectional transport in fully reconfigurable two dimensional optical ratchets, Phys. Rev. Lett. $\mathbf{118}$, 138002 (2017).
		\bibitem{Feynman} R. P. Feynman, R. B. Leighton, and M. Sands, \textit{The Feynman Lectures on Physics Vol. I} (Addison-Wesley, Reading, MA, 1963).
		\bibitem{hanggi} P. H\"{a}nggi and F. Marchesoni, Artificial Brownian motors: Controlling transport on the nanoscale, Rev. Mod. Phys. $\mathbf{81}$, 387-442 (2009).
		\bibitem{Nicollier} P. Nicollier, C. Schwemmer, F. Ruggeri, D. Widmer, X. Y. Ma, and A. W. Knoll, Nanometer-scale-resolution multichannel separation of spherical particles in a rocking ratchet with increasing barrier heights, Phys. Rev. Appl. $\mathbf{15}$, 034006 (2021).
		\bibitem{Schwemmer} C. Schwemmer, S. Fringes, U. Duerig, Y. K. Ryu, and A. W. Knoll, Experimental observation of current reversal in a rocking Brownian motor, Phys. Rev. Lett. $\mathbf{121}$, 104102 (2018).              
		
		\bibitem{Matthias} S. Matthias and F. M\"{u}ller, Asymmetric pores in a silicon membrane acting as massively parallel brownian ratchets, Nature $\mathbf{424}$, 53-57 (2003).
		\bibitem{Lout} K. Loutherback, J. Puchalla, R. H. Austin, and J. C. Sturm, Deterministic microfluidic ratchet, Phys. Rev. Lett. $\mathbf{102}$, 045301 (2009).
		\bibitem{Motegi} T. Motegi, H. Nabika, Y. Q. Fu, L. L. Chen, Y. L. Sun, J. W. Zhao, and K. Murakoshi, Effective Brownian ratchet separation by a combination of molecular filtering and a self-spreading lipid bilayer system, Langmuir $\mathbf{30}$, 7496-7501 (2014).
		\bibitem{Verleger} S. Verleger, A. Grimm, C. Kreuter, H. M. Tan, J. A. van Kan, A. Erbe, E. Scheer, and J. R. C. van der Maarel, A single-channel microparticle sieve based on Brownian ratchet, Lab Chip $\mathbf{12}$, 1238-1241 (2012).
		\bibitem{Morfill} G. E. Morfill and A. V. Ivlev, Complex plasmas: An interdisciplinary research field, Rev. Mod. Phys. $\mathbf{81}$, 1353-1404 (2009).
		
		\bibitem{Shukla} P. K. Shukla and B. Eliasson, Fundamentals of dust-plasma interactions, Rev. Mod. Phys. $\mathbf{81}$, 25-44 (2009).
		\bibitem{POPREV} J. Beckers, J. Berndt, D. Block, M. Bonitz, P. J. Bruggeman, L. Cou\"{e}del, G. L. Delzanno, Y. Feng, R. Gopalakrishnan, F. Greiner, P. Hartmann, M. Hor\'{a}nyi, H. Kersten, C. A. Knapek, U. Konopka, U. Kortshagen, E. G. Kostadinova, E. Kova\v{c}evi\'{c}, S. I. Krasheninnikov, I. Mann, D. Mariotti, L. S. Matthews, A. Melzer, and M. van de Kerkhof, Physics and applications of dusty plasmas: The Perspectives, Phys. Plasmas $\mathbf{30}$, 120601 (2023).
		\bibitem{Wang} Y. N. Wang, L. J. Hou, and X. G. Wang, Self-consistent nonlinear resonance and hysteresis of a charged microparticle in a rf sheath, Phys. Rev. Lett. $\mathbf{89}$, 155001 (2002).              
		\bibitem{Chu} J. H. Chu and L. I, Direct observation of Coulomb crystals and liquids in strongly coupled rf dusty plasmas, Phys. Rev. Lett. $\mathbf{72}$, 4009-4012 (1994).
		\bibitem{Thomas} H. M. Thomas and G. E. Morfill, Melting dynamics of a plasma crystal, Nature $\mathbf{379}$, 806-809 (1996).                                                             
		
		\bibitem{Ott} T. Ott, M. Bonitz, P. Hartmann, and Z. Donk\'{o}, Spontaneous generation of temperature anisotropy in a strongly coupled magnetized plasma, Phys. Rev. E $\mathbf{95}$, 013209 (2017).
		\bibitem{Killer} C. Killer, T. Bockwoldt, S. Sch\"{u}tt, M. Himpel, A. Melzer, and A. Piel, Phase separation of binary charged particle systems with small size disparities using a dusty plasma, Phys. Rev. Lett. $\mathbf{116}$, 115002 (2016).
		\bibitem{Du} C. R. Du, V. Nosenko, H. M. Thomas, Y. F. Lin, G. E. Morfill, and A. V. Ivlev, Slow Dynamics in a Quasi-Two-Dimensional Binary Complex Plasma, Phys. Rev. Lett. $\mathbf{123}$ 185002 (2019).
		\bibitem{Feng} D. Huang, M. Baggioli, S. Y. Lu, Z. Ma, and Y. Feng, Revealing the supercritical dynamics of dusty plasmas and their liquidlike to gaslike dynamical crossover, Phys. Rev. Res. $\mathbf{5}$ 013149 (2023).
		\bibitem{Bajaj} P. Bajaj, S. Khrapak, V. Yaroshenko, and M. Schwabe, Spatial distribution of dust density wave properties in fluid complex plasmas, Phys. Rev. E $\mathbf{105}$ 025202 (2022).
		
		\bibitem{Joshi} E. Joshi, M. Y. Pustylnik, M. H. Thoma, H. M. Thomas, and M. Schwabe, Recrystallization in string-fluid complex plasmas, Phys. Rev. Res. $\mathbf{5}$ L012030 (2023).
		\bibitem{HuangD} D. Huang, S. Y. Lu, M. S. Murillo, and Y. Feng, Origin of viscosity at individual particle level in Yukawa liquids, Phys. Rev. Res. $\mathbf{4}$ 033064 (2022).
		\bibitem{Wong} C. S. Wong, J. Goree, Z. Haralson, and B. Liu, Strongly coupled plasmas obey the fluctuation theorem for entropy production, Nat. Phys. $\mathbf{14}$, 21-24 (2018).
		\bibitem{Kalman} G. Kalman, M. Rosenberg, and H. E. DeWitt, Collective modes in strongly correlated yukawa liquids: waves in dusty plasmas, Phys. Rev. Lett. $\mathbf{84}$, 6030-6033 (2000).  
		\bibitem{He} Y. F. He, B. Q. Ai, C. X. Dai, C. Song, R. Q. Wang, W. T. Sun, F. C. Liu, and Y. Feng,  Experimental demonstration of a dusty plasma ratchet rectification and its reversal, Phys. Rev. Lett. $\mathbf{124}$, 075001 (2020).
		\bibitem{Wang1} S. X. Zhang, S. Wang, T. Y. Yao, M. Tian, W. L. Fan, F. C. Liu, and Y. F. He, Height-modulating horizontal transport of dust particles in a dusty plasma ratchet, Plasma Sources Sci. Technol. $\mathbf{33}$, 055008 (2024).
       \bibitem{SM} See Supplemental Material at [URL will be inserted by publisher] for Videos of positive-negative [Fig. 2(a)],  positive-positive [Fig. 2(c)], and negative-negative [Fig. 2(e)] flows of be-dispersed dust particles in experiments.
         \bibitem{COMSOL} COMSOL Multiphysics version 5.3a, www.comsol.com.  
      \bibitem{Tian} M. Tian, S. P. Li, T. Y. Yao, X. Z. Wang, F. C. Liu, and Y. F. He, Separation of micron-sized dust particles in low-pressure air using a dusty plasma ratchet, Plasma Sci. Technol. $\mathbf{27}$, 054004 (2025).
		
		
		
		
		%
		
	\end{thebibliography}
\end{document}